\documentclass[a4paper,fleqn,usenatbib]{mnras}

\usepackage{url,ulem,times,graphicx,amsmath,amsfonts,amssymb,color,epsfig,epstopdf}
\usepackage{graphics}
\usepackage{epsf}
\usepackage{bm}
\usepackage{color}
\usepackage{rotating}
\usepackage[T1]{fontenc}
\usepackage{ae,aecompl}
\usepackage{array,multirow}
\usepackage[percent]{overpic}
\usepackage{amsfonts,amssymb,amsmath}
\usepackage{aas_macros}
\usepackage{varioref,textcomp}

\usepackage{hyperref}
\hypersetup{
   colorlinks=true,
   urlcolor=blue,
   citecolor=blue,
   linkcolor=blue,
   pdfborder= 0 0 0
}

\def\msun{\,{\rm M}_\odot/h}
\def\mpc{\,{\rm Mpc}}
\def\mpch{\,{\rm Mpc}/h}

\newcommand{\blue}{\textcolor{blue}}

\def\e{\,{\boldsymbol{e_1}}}
\def\ee{\,{\boldsymbol{e_2}}}
\def\eee{\,{\boldsymbol{e_3}}}

\newcommand{\Rmnum}[1]{\uppercase\expandafter{\romannumeral #1}}

\title[Halo spin and large scale structure]
      {The build up of the correlation between halo spin and the large scale structure}
\author[P. Wang \& X. Kang]
{Peng Wang$^{1,2}$,
Xi Kang$^{1}$ \thanks{E-mail:kangxi@pmo.ac.cn}\\
$^1$Purple Mountain Observatory, the Partner Group of MPI f\"ur Astronomie, 2 West Beijing Road, Nanjing 210008, China \\
$^2$Graduate School, University of the Chinese Academy of Science, 19A, Yuquan Road, Beijing 100049, China \\
}

\date{Accepted 2017 September 20. Received 2017 September 20; in original form 2017 July 14.}

\pubyear{2017}

\begin{document}
\label{firstpage}
\pagerange{\pageref{firstpage}--\pageref{lastpage}}
\maketitle

\begin{abstract}

Both simulations and observations have confirmed that the spin of haloes/galaxies is correlated with the large scale structure (LSS) with a mass dependence such that the spin of low-mass haloes/galaxies tend to be parallel with the LSS, while that of massive haloes/galaxies tend to be perpendicular with the LSS. It is still unclear how this mass dependence is built up over time. We use N-body simulations to trace the evolution of the halo spin-LSS correlation and find that at early times the spin of all halo progenitors is parallel with the LSS. As time goes on, mass collapsing around massive halo is more isotropic, especially the recent mass accretion along the slowest collapsing direction is significant and it brings the halo spin to be perpendicular with the LSS. Adopting the $fractional$ $anisotropy$ (FA) parameter to describe the degree of anisotropy of the large-scale environment, we find that the spin-LSS correlation is a strong function of the environment such that a higher FA (more anisotropic environment) leads to an aligned signal, and a lower anisotropy leads to a misaligned signal. In general, our results show that the spin-LSS correlation is a combined consequence of  mass flow and halo growth within the cosmic web. Our predicted environmental dependence between spin and large-scale structure can be further tested using galaxy surveys.

\end{abstract}

\begin{keywords}
methods: numerical ---
methods: statistical ---
galaxies: haloes ---
Galaxy: halo ---
cosmology: dark matter
\end{keywords}



\section{Introduction}
\label{sec:intro}

In the currently favoured model for structure formation, initial seeds of
perturbation  in  the early  universe   were  amplified  by  gravitational
instability and dark matter haloes were  formed in the regions where the
linear density contrast  reach some threshold \citep{1972ApJ...176....1G}. On
large scales, the  matter distribution is characterized  by the cosmic web
\citep{1996ApJS..103....1B}.  As dark  matter haloes are residing  in different
cosmic  environments  (voids,  sheets,  filaments, clusters)  and  mass  is continuously flowing out of voids, going to sheets, filaments and finally into clusters
\citep{1970A&A.....5...84Z},  it is strongly  expected that the halo  properties are
closely related to their cosmic environment or the nearby large-scale structure
(LSS).  Numerous studies, especially  those using N-body simulations,
have confirmed a few correlations between  halo properties and the LSS
\citep[e.g.,][]{2007MNRAS.381...41H, 2007MNRAS.375..489H}.

Among those  various kinds  of correlations, the  halo spin-LSS  is of
great interest since it is a result of halo formation in the tidal field
exerted  by the  mass  on large  scales, and  this  correlation can  be
directly    predicted    by    the    linear    perturbation    theory
\citep[e.g.,][]{1951pca..conf..195H, 1969ApJ...155..393P,
  1970Ap......6..320D, 1984ApJ...286...38W,  1987ApJ...319..575B}. The
tidal torque  theory (TTT) predicts that the halo spin has a  tendency to
align     with     the     intermediate    axis     of     the     LSS
\citep[e.g.,][]{1996ApJS..103....1B,
  1996MNRAS.281...84V,1993ApJ...418..544V}  under the  assumption that 
  the halo inertia  tensor is  not correlated with  the LSS.   Although this
assumption is  not always  true \cite[][]{2000ApJ...532L...5L},  it is
generally found from  simulations that the halo spin does have  a tendency to
align     with     the     intermediate    axis     of     the     LSS
\citep{2007ApJ...655L...5A, 2010MNRAS.405..274H, 2012MNRAS.421L.137L, 2016ApJ...825...49C},
       consistent        with      the        TTT       predictions
\citep[e.g.,][]{2004MNRAS.349.1039N}.

Although the predicted correlation  between halo spin and the intermediate axis of
the LSS is confirmed by  N-body simulations, a comparison with the observational  data is not trivial. That is because the middle axis of LSS in a
filamentary structure is not prominent and thus being difficult to be identified.
As  the  mass along  the  least compressed direction (later  labelled as $\eee$)
is  relatively easy to be identified, most  studies have focused on  the
spin-e3 correlation.  \cite{2007ApJ...655L...5A}  firstly   reported  that,
although  this correlation is  weak, the  signal is highly  confident and
depends on halo  mass. The  spin  of low-mass  haloes have  a  tendency to
align (parallel) with the filament, and that of high-mass haloes are prone to be
perpendicular with  the filament.   This mass  dependence and  its redshift
evolution are later confirmed by many studies which used N-body and hydro-dynamical
simulations      \citep[][]{2007MNRAS.381...41H, 2009ApJ...706..747Z,
2012MNRAS.427.3320C,    2013ApJ...762...72T, 2013ApJ...766L..15L,
2014MNRAS.443.1090F,    2014MNRAS.445L..46W,
2014MNRAS.444.1453D,2015MNRAS.446.2744L,       2017MNRAS.468L.123W}.  Indeed,
such  a spin-LSS correlation  with a galaxy  mass (morphology) dependence  is
also found  in  the  observational  data of  the  SDSS
\citep{2013ApJ...775L..42T}.

The mass  dependence of the  spin-e3 correlation  or the flip  of this
relation  can  be  understood  by studying  the  halo  mass  accretion
history, as halo spin is a  result of  mergers and conservation  of angular
momentum   \citep[e.g.,][]{2004MNRAS.352..376A,  2005ApJ...627..647B}.
It  was  previously found  from  simulations  that most  subhaloes  are
accreted   along   filaments  \citep[e.g.,][]{2005MNRAS.364..424W,
  2014MNRAS.443.1274L, 2014ApJ...786....8W,  2015ApJ...807...37S}, but
a universal direction of mass accretion is unable to explain the flip
of \blue{the} spin-e3 correlation for  low-mass  haloes.   \cite{2015ApJ...813....6K}  used
high-resolution  simulations  to  resolve   the  formation  history  of
low-mass  haloes and  they found  that  mass accretion  is indeed  not
universal,  and  the accretion  of  subhaloes  in low-mass  haloes  are
perpendicular to the filament.   This non-universal mass accretion can
well  explain the  flip of the spin-e3  relation. In  addition, the  mass
accretion pattern  can be understood  in a wider context  of mass
flow within the cosmic  web. By tracing the mass flow  in different cosmic
environments,    a    few    studies    \citep[][]{2012MNRAS.427.3320C,
  2014MNRAS.445L..46W, 2015MNRAS.446.2744L} pointed  out that low-mass
haloes are mainly formed by smoothing  accretion through the wind  of flows
embedded in misaligned walls, and  massive haloes are products of major
mergers in filaments. \cite{2017MNRAS.468L.123W}  further find that the
spin-e3 correlation  is closely related  to halo formation time  and the
transition time when the halo environment changes.

Although those above studies provide a good explanation for the mass dependence of 
halo  spin-e3 correlation,  it is  not clear  how this  correlation is
built  up  in  detail.  For  example,  for  massive  haloes  with the spin
perpendicular to the filament,  if we trace them back  to higher redshift,
should  we see  a parallel  signal between  the progenitors'  spin and the filament? How is this spin flip related to the halo mass growth history
and the velocity of accreted  subhaloes?  Other  than these questions,  we also
want  to  investigate  if  the  spin-e3  correlation  depends  on  the
properties of the cosmic web, such  as the degree of anisotropy of the
LSS.  These are  the  main  motivations of  this  work.  Our paper  is
organized   as   follows.   Sec.~\ref{sec:simulation}   presents   the
simulation data, the method to  quantify the halo environment and the halo
spin-e3 correlation.  In Sec.~\ref{sec:results} we show the results in
detail  including  the effects  of  smoothing  length, the evolution  of  the
spin-e3 relation,  the dependence  of the  spin-e3 correlation  on the
anisotropy  of the  cosmic  web.  Conclusions and Discussion are presented in Sec.~\ref{sec:con}.


\section{Simulation and Method}
\label{sec:simulation}

The simulation data used in this work  is the same as that used in our
two                           previous                          papers
\citep[][]{2015ApJ...813....6K,2017MNRAS.468L.123W}.   In   \cite{2015ApJ...813....6K}  we study the accretion of halo mass and its
relation with halo major axis and the large-scale environment. In the second
paper, we investigate the spin-LSS  correlation with dependence on halo
formation time and the transition  time when the halo environment changes. Readers  interested  in  these  studies  can  refer  to  those  two
papers. In the  following, we shortly introduce the data  and method to
classify halo environment.

The   N-body   simulation   was    run   using   the   GADGET-2   code
\citep[][]{2005MNRAS.364.1105S} and it follows $1024^{3}$ particles in
a periodic  box of $200  \mpch$ with cosmological parameters  from the
$WMAP7$   data   \citep[][]{2011ApJS..192...18K}   namely:
$\Omega_{\Lambda}=0.73$,      $\Omega_{m}=0.27$,      $h=0.7$      and
$\sigma_{8}=0.81$.  The  particle mass  in this simulation  is 
$5.5\times 10^{8}\msun$ and  60 snapshots are stored  from redshift 10
to 0.   We identify dark matter  haloes from the simulation  using the
standard          friends-of-friends          (FOF)          algorithm
\citep[][]{1985ApJ...292..371D} with  a linking length that  is 0.2 times
the mean  inter-particle separation. For  each FOF halo,  we determine
its virial  radius, defined as the  radius centred on the  most bound
particle inside  of which the  average density is $200$ times 
the average density  of the universe \citep{1998ApJ...495...80B}. The mass
inside the virial  radius is called the virial mass  $M_{vir}$. We use the
SUBFIND   algorithm    \citep[][]{2001MNRAS.328..726S}   to   identify
subhaloes  in each  FOF halo,  and construct  their merger  trees 
\citep[see][for details]{2005ApJ...631...21K}. Usually, each halo has more than one progenitor at earlier times, and we refer to the most massive one as the main progenitor. By tracing the main progenitor back in time we can obtain the main formation history of the halo. We will use the merger trees to study how the spin-LSS correlation is built up with time in Sec.~\ref{sec:evolution}.

\begin{figure*}
\centerline{\epsfig{figure=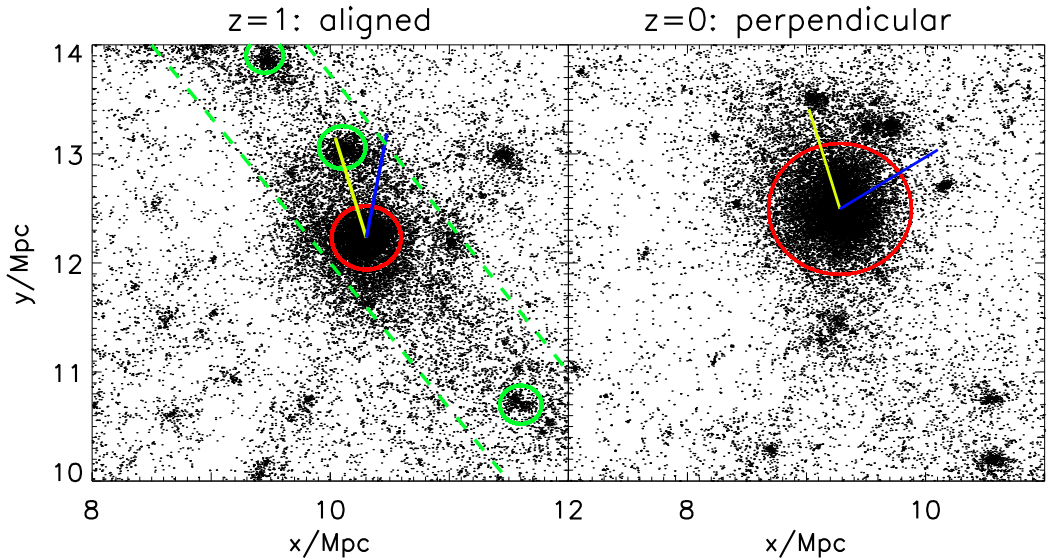,width=0.85\textwidth}}
\caption{Left: the 2-D mass distribution around a host halo (with mass $\sim
5\times10^{11}\msun$) at redshift $z=1$. The red circle denotes its virial
radius, and cyan circles denote those subhaloes (haloes) that will merge with the
host halo in a short time. The region between the two cyan dashed lines indicates a
filamentary structure around the host halo.  Right: 2-D slice at redshift $z=0$ around the descendant halo. The halo (with mass $\sim3\times10^{13}\msun$) is the
descendant of those shown in the left panel. In this example, the LSS environment of the host halo changes from filament to cluster.}
\label{fig:slice}
\end{figure*}

The spin $J$ of a halo is measured as, 
\begin{equation}
  \boldsymbol{J}=\sum_{i=1}^N m_i \boldsymbol{r_i} \times (\boldsymbol{v_i}-\boldsymbol{\bar{v}}),
\label{equ:am}
\end{equation}
where  $m_i$  is the particle  mass,  $\boldsymbol{r_i}$  is the  position
vector of particle $\boldsymbol{i}$ relative  to the halo centre, which
is defined  as the position  of the most  bound particle in  the halo.
$\boldsymbol{v_i}$   is   the   velocity    of particle
$\boldsymbol{i}$, and  $\boldsymbol{\bar{v}}$ is  the mean  velocity of
all halo particles.

To determine the large scale environment of each halo, we employ
the method of Hessian matrix used in many previous works
\citep[e.g.][]{2007A&A...474..315A, 2009ApJ...706..747Z,
2015ApJ...813....6K}. The Hessian matrix of the smoothed
density field at the halo position is defined as:
\begin{equation}
  H_{ij}=\frac {\partial^2\rho_s(\boldsymbol{x})} 
  {\partial x_{j} \partial x_{j}},
\label{equ:hessian}
\end{equation}
where  $\rho_s(\boldsymbol{x})$ is  the  smoothed  density field.  The
eigenvalues       of       the Hessian       matrix       are       sorted
($\lambda_{1}<\lambda_{2}<\lambda_{3}$),  and   the  eigenvectors  are
marked  as  $\e$, $\ee$  and  $\eee$,  respectively. According  to  the
Zel'dovich    approximation    \citep[][]{1970A&A.....5...84Z},    the
eigenvalues of the  tidal field can be used to  define the large scale
environment  of  each  dark  matter  halo.   The  number  of  positive
eigenvalues  is used  to  classify  the  environments of  a  halo as \citep[][]{2007MNRAS.381...41H,       2007MNRAS.375..489H,
  2009ApJ...706..747Z},

\begin{itemize}
  \item $cluster$:  no positive eigenvalue;
  \item $filament$: one positive eigenvalue;
  \item $sheet$: two positive eigenvalues;
  \item $void$: three positive eigenvalues;
\end{itemize}

In  N-body  simulations,  the  density field  is  discontinuous, 
so it  is necessary  to smooth  the particle
distributions. The  commonly used  way is to  divide the  simulation box
into a number  of uniform grid points, then  each particle is assigned
to several nearby grid points  in an appropriate way, mostly using the
Cloud-in-Cell interpolation. The density field can be then obtained by
smoothing  each  grid  by  using  a  Gaussian  filter  $G_{R_s}$  with 
a smoothing scale  $R_s$.

The smoothing scale $R_s$ is not  specified by the theory directly but
depends on the objects  one wants to study. For example,  if we want to
obtain the  large scale  environment of  a dark  matter halo  with virial
radius $\sim$  $100$ $\rm kpc$,  the smoothing scale must  be slightly
larger than the  physical size of this dark matter  halo, around a few
hundreds  kpc.  However,  it  can  not be  too  large,  otherwise  the
surrounding environment will  be affected by the  distribution of mass
on a very  larger scale, which  is not physically  linked to the  halo in
question. In  most works \cite[e.g.,][]{2007A&A...474..315A,2007MNRAS.381...41H,
2007MNRAS.375..489H, 2009ApJ...706..747Z, 2012MNRAS.427.3320C, 2013ApJ...762...72T} they used  a  constant
smoothing length of  $2 \rm Mpc$. For more details on how to properly select the smoothing
length, see  previous  works  \citep[e.g.,][]{2007ApJ...655L...5A, 2007MNRAS.381...41H, 2014MNRAS.443.1090F}.

In  this work,  we use  two kinds  of smoothing  methods.  The  first one
follows that used in \cite{2015ApJ...813....6K} in which $R_{s}=X\mpc/(1+z)$ (Method \Rmnum{1}),
where $X$ is $2$ or $5$. This smoothing length is the same for all haloes but is 
a function of redshift, as  it is  more  physically  reasonable and  can  better describe  the
evolution of  the large scale  environment. \cite{2012MNRAS.421L.137L}, 
instead,  suggested that a more  physical smoothing length should depend on
the  mass of  the  halo and  is  related to  the  halo virial  radius,
$R_{s}=X  R_{vir}$ (Method \Rmnum{2}), where  $X$ is  between $4$  and $8$. We will later show that our results are not strongly affected by the selection of the smoothing lengths.

In some  previous works on the  correlation between halo spin  and the
LSS \citep[e.g.,][]{2007ApJ...655L...5A}, the direction  of the LSS is
chosen based on  the geometric feature of the  structure. For example,
in a  filament region the  LSS is defined  as the direction  along the
filamentary structure,  corresponding to  $\eee$,  and in  a sheet
environment, the direction is chosen  to be perpendicular to the sheet
plane (the $\ee-\eee$ plane), so  along the $\e$ direction. As pointed
out by  \cite{2014MNRAS.443.1274L} and  \cite{2015ApJ...813....6K}, $\eee$
is  a good  and universal  definition of  the direction  of LSS.   For
instance,  in  a filament  region,  $\eee$ along  the  filamentary
structure, and  in a cluster  region, $\eee$ is the  latest collapsed,
least compressed direction. It is worth noting that when we talk
about  any  correlations  with  the  large-scale  environment,  it  is
important to have a consistent  definition of the environment. In this
work, we  thus use the $\eee$  as the universal definition  of the LSS
and use it to specify the correlation between halo spin and the LSS.

The correlation between  halo spin and the $\eee$  is quantified using
their alignment angle, $\cos\theta_3$, as,
\begin{equation}
  \cos\theta_{3}=\boldsymbol{j} \cdot \eee,
\end{equation}
where $\boldsymbol{j}$ is the spin vector and $\eee$ is the
eigenvector calculated by Equation~\ref{equ:hessian} at the position
of the halo. If the halo spin is randomly oriented in 3-D space relative
to $\eee$, the expected value of $<\cos(\theta_3)>$ is 0.5. In case
$<\cos(\theta_3)>$ is larger than 0.5, we refer it to an alignment
(parallel) between the spin and the LSS, and for $<\cos(\theta_3)>$ lower
than 0.5, we call it a mis-alignment (perpendicular) between the spin and
the LSS.


\section{Results}
\label{sec:results}

In this section, we first show the spin-$\eee$  correlation with its
dependence on  smoothing lengths, halo  mass and redshift. Then,  we
trace the haloes  back in time to see how  the spin-$\eee$ correlation
is built  up for haloes  with different  masses, and investigate  how the
flip of the  relation is related to their anisotropic  mass growth. We
will also  show that  the anisotropy  of the  halo environment  can be
quantified using a simple parameter,  and it also partly determines the degree
of the halo spin-LSS alignment.

\subsection{Spin flips}
\label{sec:flips}

In order to have an intuition about the  correlation between halo spin with the
LSS and  its evolution, in  Fig.~\ref{fig:slice} we show the  2-D mass
distribution  around  a   dark  matter  halo  whose   virial  mass  is
$\sim3\times10^{13}\msun$ at redshift $z=0$, and its main progenitor is
$\sim5\times10^{11}\msun$ at redshift $z=1$.  The left panel shows the
progenitors at  $z=1$, where  the main progenitor  is marked  with a red
circle and the other progenitors (for  clarity, only three of them are
shown)  are  marked as  green  circles.   The  yellow line  gives  the
direction  of   the  LSS  ($\eee$),  which   resembles  a  filamentary
structure,  and  the  blue  line   indicates  the  spin  of  the  main
progenitor.  In this  case the halo spin is aligned  with the LSS. The
right  panel  shows  that  after   mergers  the  final  halo  spin  is
misaligned  (or perpendicular)  with the  direction of  the LSS,  for
which  the  environment  is  turned   into  a  cluster  region.   This
particular  case shows  that in  a filamentary  structure, the  mass is
mainly  accreted along  the filament,  and after  most of the mass along the filament has collapsed (merged into the main halo), and the LSS environment
will  possibly evolve  into a  cluster region,  in which  the spin  is
perpendicular to the recently collapsed direction.

Previous     studies    \citep[][]{2007ApJ...655L...5A,
  2007MNRAS.381...41H}  usually investigated  the spin-LSS  correlations for
haloes  in sheet  and filament  environments separately.  These studies
found that the spin of haloes in sheet  tends to lie in the sheet plane, and
in filaments  there is a  transition halo mass  around $10^{12}M_{\odot}$,
below which  halo spin is  parallel to filaments and above  which the halo
spin  is perpendicular  to the filament. \cite{2017MNRAS.468L.123W}  find
that, similar  to the  case for the filament,  the spin-e3  correlation of
haloes  in cluster  environment  also  has a  mass  dependence, and  the
transition mass is slightly smaller than that in filament. In this work we
investigate the spin-LSS  correlation for haloes as a whole,  and we do
not consider halo environment, but just focus on the mass dependence.

\begin{figure}
\centerline{\epsfig{figure=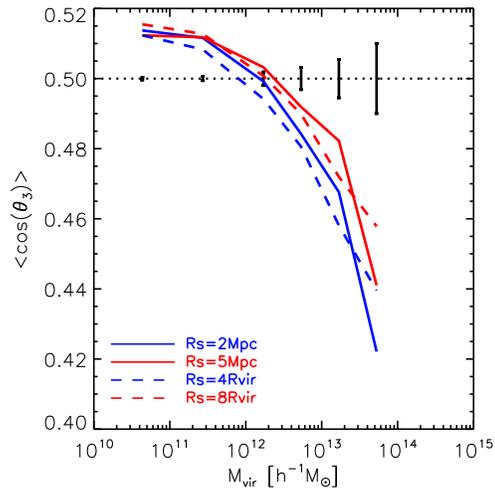,width=0.4\textwidth}}
\caption{The halo spin-$\eee$ correlation at $z=0$ as a function of halo virial mass. The solid (Method \Rmnum{1}) and dashed (Method \Rmnum{2}) lines are for two different smoothing methods. The horizontal dashed line corresponds to a random distribution between halo spin and $\eee$ of the large scale structure. It is seen that the alignment depends on halo mass, and the flip of this correlation is at around $10^{12}M_{\odot}$ with a slight dependence on the smoothing length. The error bars are calculated by $1/\sqrt{N}$, in which N is the number in each bin. The same method is adopted in following figures.}
\label{fig:spin_flips}
\end{figure}

In Fig.~\ref{fig:spin_flips} we show  the spin-e3 correlation at $z=0$
with  two different  smoothing methods.  In agreement  with a  previously
similar study by \cite{2014MNRAS.443.1090F}, we find that, without
considering halo  environment, the spin-e3 correlation  is similar to
that of haloes  in filaments, with a transition mass  at which the spin-e3
correlation  flips.  Note  that although  the  alignment/mis-alignment
signal  is  very  weak,  it   departs  significantly  from  a  random
distribution, which  can be seen  from the errorbars on  the horizontal
dash   line.  Fig.~\ref{fig:spin_flips}   also  shows   that  adopting
different smoothing methods will produce a slightly different transition mass,
ranging within a  factor of 2 between $7  \times 10^{11}M_{\odot} \sim
2\times 10^{12}M_{\odot}$.  This transition mass with a weak dependence on
smoothing scales  is in  broad agreement with  previous results
\citep[][]{2007ApJ...655L...5A,                   2014MNRAS.440L..46A,
  2012MNRAS.427.3320C,  2016IAUS..308..421P, 2017MNRAS.468L.123W}.  As
the effects  of the smoothing length are  very weak, in the  following we will
only show results  from the second smoothing methods with $R_{s}=4R_{vir}$.  We have tested that
our results  and conclusions  are not affected  by using  other smoothing
lengths.

\begin{figure*}
\centerline{\epsfig{figure=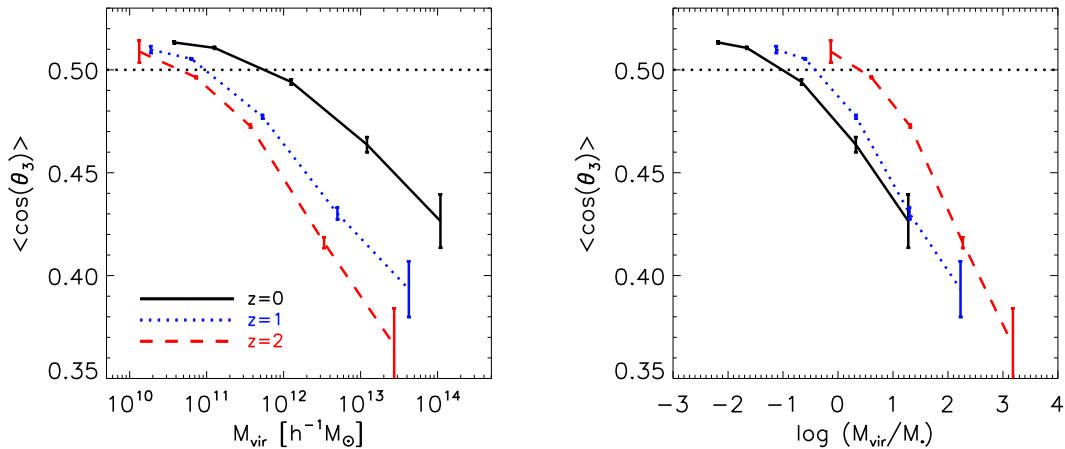,width=0.85\textwidth}}
\caption{Left: the halo spin-e3 alignment at different redshifts.  Right: as the left panel, but the halo mass is normalized by $M_{\star}$, the characteristic mass of halo collapsing at a given redshift. The two panels show different redshift dependence,  while the right panel reveals a real evolution pattern at given normalized halo mass (see text for more detail).} 

\label{fig:spin_Mass_Mstar}
\end{figure*}

In  Fig.~\ref{fig:spin_Mass_Mstar}  we  show  the  mass  and  redshift
dependence of the  spin-e3 correlations. Note that in  the right panel
the  halo  mass is  normalized  by  the  characteristic mass  of  halo
collapsing at  a given redshift.  The left panel shows  that the
mass dependence of  spin-e3 correlation is seen at  all redshifts, but
the   transition  mass   where   the   correlation  flip   from
perpendicular to parallel  is lower at high redshifts.   At given halo
mass, the alignment signal is  lower at high redshift, indicating that
the halo spin  is more likely to  be perpendicular to the  LSS at high
redshift.  However, this  is not an evolution pattern  for the spin-e3
relation.   As the  halo mass  also  grows with  redshift, so  we
  should not  inspect the  relation of  the relation  at a  given halo
  mass. More  physically interesting  results are  seen in  the right
panel  of  Fig.~\ref{fig:spin_Mass_Mstar},  where  the  halo  mass  is
normalized  by  the characteristic  mass,  $M_{\star}$,  which is  the
typical  mass  of  halo  collapsing  at a  given  redshift.  The  mass
$M_{\star}$ can  be calculated  using the  matter power  spectrum with
a given  set  of  cosmological  parameters  and  a  given
redshift  \citep[see][for  more details]{2013ApJ...762...72T}.  It  is
seen that for haloes with  a given $M_{vir}/M_{\star}$, i.e., the same
stages of collapse  (\cite{2013ApJ...762...72T}), the alignment signal
is higher  at high  redshift, showing that  during the  evolution halo
spin is more likely to be parallel to the LSS at early times.  As time
passes by, the alignment becomes weaker and the spin is inclined to be
perpendicular to the LSS at  low redshift.  These redshift dependencies are     in      broad     agreement     with      previous     studies
\citep{2007MNRAS.381...41H, 2013ApJ...762...72T}.


\subsection{Spin evolution and mass accretion anisotropy}
\label{sec:evolution}

\begin{figure*}
\centerline{\epsfig{figure=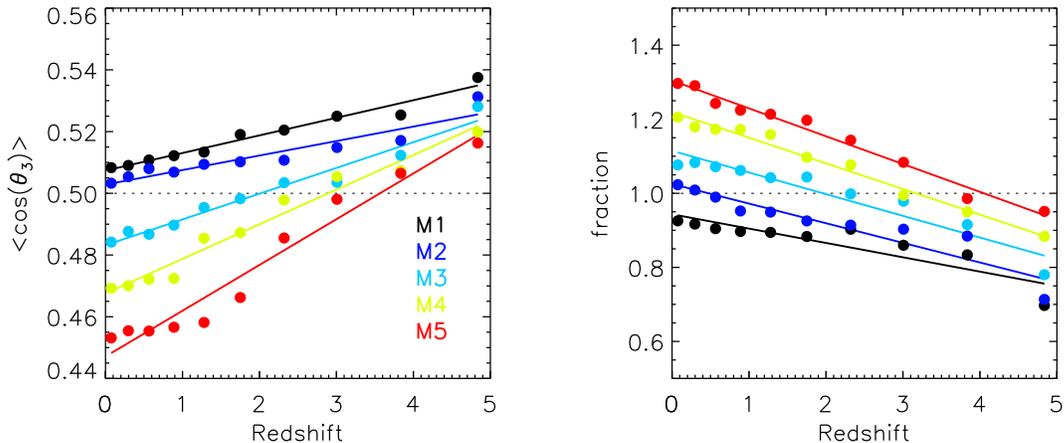,width=0.85\textwidth}}
\caption{Left: the evolution of the mean cosine of the angle, $<\theta_{3}>$, between the spin of halo's progenitors and the $\eee$ at corresponding redshift. The descendant haloes at $z=0$ are divided into 5 mass bins, marked from $\rm M1$ to $\rm M5$ with increasing mass (see Table.~\ref{table:samples}). Right: the evolution of the anisotropic mass accretion for haloes in different mass bins. Here $fraction$ is the ratio between the cumulative mass accreted along $\eee$ ($0^{\circ} < \theta < 45^{\circ}$) and the cumulative accreted mass perpendicular to $\eee$ ($45^{\circ} < \theta < 90^{\circ}$) at a given redshift bin. The points are the measurements from simulations and the lines are the linear fits to the data, with parameters listed in Table.~\ref{table:fitting}}
\label{fig:SpinLSS_MassFlow_Evolution}
\end{figure*}

\begin{table}
\caption{We divide haloes at $z=0$ into five mass bins, from $\rm M1$ to $\rm M5$ with increasing mass. The table lists the range of mass and number of haloes in each mass bin.}
\begin{center}
\begin{tabular}{ccc}
\hline
sample   &        Mass range ($z=0$)    & Num. of haloes \\
\hline
  M1     & $[10^{10}-10^{11}] \msun$    &   570123       \\
  M2     & $[10^{11}-10^{12}] \msun$    &   174818       \\
  M3     & $[10^{12}-10^{12.5}] \msun$  &   17351        \\
  M4     & $[10^{12.5}-10^{13}] \msun$  &   6207         \\
  M5     & $>10^{13} \msun$             &   2830         \\
\hline
\end{tabular}
\end{center}
\label{table:samples}
\end{table}

\begin{table*}
\caption{The parameters obtained from a linear fitting, $y=a*redshift+b$, are listed
below.  Slopes, zero-points and their errors are signed with $a$, $b$ and
$\sigma$, with subscript $\theta$ for the left panel in
Fig.~\ref{fig:SpinLSS_MassFlow_Evolution} and subscript $f$ for the right panel.}
\begin{center}
\begin{tabular}{ccccccccc}
\hline
sample & $a_\theta$ & $b_\theta$ & $\sigma_{a_\theta}$ & $\sigma_{b_\theta}$ & $a_f$ & $b_f$ & $\sigma_{a_f}$ & $\sigma_{b_f}$ \\
\hline
M1 & 0.0057008 & 0.507353 & 0.151267 & 0.343309 & -0.0387258 & 0.94338 & 0.187814 & 0.559281 \\
M2 & 0.0046992 & 0.502829 & 0.150392 & 0.349159 & -0.0530214 & 1.02523 & 0.193150 & 0.540680 \\
M3 & 0.0083454 & 0.483183 & 0.148909 & 0.354379 & -0.0585716 & 1.11478 & 0.200449 & 0.515152 \\
M4 & 0.0112145 & 0.467537 & 0.147422 & 0.360299 & -0.0693331 & 1.21954 & 0.210341 & 0.495142 \\
M5 & 0.0148303 & 0.447117 & 0.146150 & 0.361961 & -0.0751330 & 1.30434 & 0.217498 & 0.476506 \\
\hline
\end{tabular}
\end{center}
\label{table:fitting}
\end{table*}

In order to understand in detail how the  spin-e3 correlation is built up and
how it is related to halo mass  growth, we trace haloes back in time
to study  the spin-e3 correlation  of their main progenitors  with the
large scale structure  at different redshifts. We  select haloes within
different mass bins at $z=0$,  and in Tab.~\ref{table:samples} we list
the mass ranges and  number of haloes in each mass  bins. For each dark
matter halo we trace its main  progenitor back to higher redshifts, and we
also obtain the  $\eee$ direction  of the LSS  around each  main progenitor
using the Hessian matrix method.

In the left panel of Fig.~\ref{fig:SpinLSS_MassFlow_Evolution}, we show
the evolution of $\cos\theta_3$ between the  spin and $\eee$
for the main progenitors of haloes selected in five different mass bins
at $z=0$, where the halo mass is increasing from the top to the bottom
lines.  In the  right panel, we show the evolution  of the anisotropic
mass  accretion history  of  those haloes.  Here  the anisotropic  mass
accretion parameter  $fraction$ is  defined as  the ratio  between the
total mass accreted at each redshift by the  main progenitor along the $\eee$ direction
(within  an  angle between  $0^{\circ}$  and  $45^{\circ}$), and  those
accreted perpendicular to $\eee$, within an angle between $45^{\circ}$
and $90^{\circ}$.  In  both panels the lines are the  best linear fits to the
points with fitted parameters given in Table.~\ref{table:fitting}.

The left panel shows that there is a strong evolution of the alignment between halo spin and $\eee$. For massive haloes (red points), their spin is perpendicular to the LSS at lower redshift, but for their main progenitors at high redshifts, such as at $z>3$, the halo spin is parallel to $\eee$. For low-mass haloes (black points), their spin is parallel to $\eee$ all the time, but the alignment is much better at higher redshift. Of course here the signal is for the average, and it is definitely true that for a single halo its spin evolution can be much more complicated. This plot shows that, on average, at high redshifts the halo spin is more likely to be parallel to the LSS, but with the halo mass growth, the halo spin will more likely evolve into being perpendicular to the LSS, and this evolution is faster for massive haloes. As pointed out in \cite{2017MNRAS.468L.123W}, this rapid evolution for massive haloes is related to their migration time that massive haloes enter filament earlier, so the mass flow along the filament will lead to a quick spin flip. 

The evolution of the halo spin-$\eee$ correlation is strongly related to the anisotropic mass accretion history, which is shown in the right panel of Fig.~\ref{fig:SpinLSS_MassFlow_Evolution}. 
It shows that, at the early time, the mass accretion in all progenitors is mainly perpendicular to the $\eee$,  and their evolution is dependent on halo mass. For low-mass haloes, the parameter $fraction$ is lower than 1 across the time, but for massive haloes, the mass accretion proceeds to be along the $\eee$ at lower redshifts. Under the natural assumption that the orbital angular momentum of accreted mass will be converted to halo spin \citep[][]{2014MNRAS.445L..46W}, it is easy to understand that once the mass accretion along $\eee$ is dominated, the final halo spin will be perpendicular to $\eee$, in good agreement with previous work \citep[][]{2014MNRAS.443.1274L, 2015ApJ...813....6K}. The transition redshifts for the flip of the spin-$\eee$ correlation in the left panel correspond well with those in the right panel for different halo mass bins, showing that the halo spin is strongly correlated with the mass accretion history.

\begin{figure*}
\centerline{\epsfig{figure=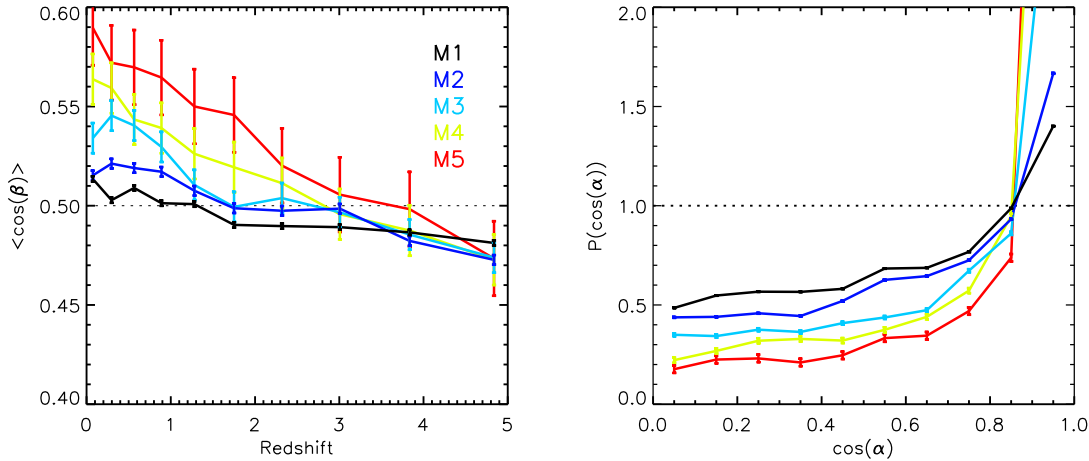,width=0.85\textwidth}}
\caption{
Left panel: the mean $\cos\beta$ as a function of redshift, in  which $\beta$ is the angle between the
halo velocity vector and the direction of LSS, $\eee$.  Right
panel: the distribution of the angle between the $\eee$ at $z=z_i$ and the $\eee$ at $z=0$. It shows that the $\eee$ is better aligned during the halo evolution, especially for massive halos.}
\label{fig:host_vel_e3_swing}
\end{figure*}

On larger scales, the motion of dark matter haloes should follow the mass flow within the cosmic web.  
\cite{1991QJRAS..32...85I} suggested that mass and velocity flows are transported in such a way: matter flows out from the void into the sheet plane and then flows inside the sheet plane to the intersection of sheets where the filament is formed. Inside the filament, the mass is mainly flowing along the filamentary structure, and finally into a cluster region where the filaments intersect with each other.  Such a model of mass transport within cosmic web is clearly illustrated using simulations. (\cite{2014MNRAS.441.2923C}, their Fig.35). In the left panel of Fig.~\ref{fig:host_vel_e3_swing} we show the evolution of the average angle $<\cos\beta>$ between the halo velocity and the LSS $\eee$. The plot shows that, at high redshift, the halo velocity is slightly perpendicular to $\eee$, but becomes more parallel to $\eee$ at lower redshift. At early times most haloes are in the sheet environment, where the mass flow is mainly from voids and thus being perpendicular to the sheet plane. As time goes on, halo environment is progressively changing into filament or cluster, so the halo velocity is mainly along $\eee$. Both the mass and velocity flows well explain the evolution of spin-$\eee$ relation as seen in Fig.~\ref{fig:SpinLSS_MassFlow_Evolution}.

Although the halo environment is evolving with time, our definition of $\eee$ is the slowest collapsed direction, thus its direction is not expected to change much during the halo evolution. To check this assumption, in the right panel of Fig.~\ref{fig:host_vel_e3_swing} we show the distribution of the angle,$\alpha$, between the LSS $\eee$ around the main progenitor at $z>0$ with the $\eee$ of the descendant halo at $z=0$. Similar with previous results, the $z=0$ descendant haloes are selected in different mass bins. The panel shows that the distribution is skewed to higher $\cos(\alpha)$, indicating that indeed the $\eee$ direction does not change much during the evolution. It is also seen that for massive haloes the $\eee$ retains its stability much better.

\subsection{FA dependence}
\label{sec:fa}

\begin{figure}
\centerline{\epsfig{figure=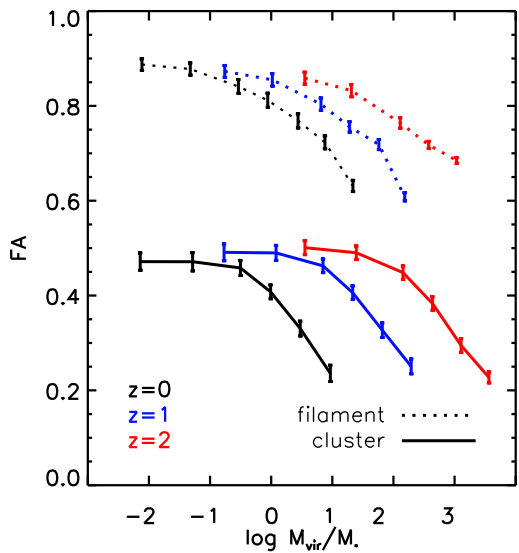,width=0.45\textwidth}}
\caption{The fractional anisotropic ($\rm FA$) as a function of halo mass for haloes in different environments and redshifts. $\rm FA$ denotes the degree of anisotropy of the large-scale tidal field, where a lower value means that the mass along all the three directions of the tidal field has collapsed, as the case for the cluster. A higher $\rm FA$ means that the mass collapse is highly anisotropic, as the case for a sheet or filament environment.}
\label{fig:fa}
\end{figure}

The above results show that the halo spin-$\eee$ correlation depends on the anisotropic mass accretion on large scales. Usually, for haloes in a given cosmic environment, the anisotropic collapse along the eigenvectors of the tidal field might be different. It would be interesting to see if the spin-$\eee$ correlation is also dependent on the degree of the anisotropic tidal field. \cite{2013MNRAS.428.2489L} defined an useful parameter, $fractional$ $anisotropy$ (FA), to quantify the degree of anisotropy of the tidal field:  
\begin{equation}
  {\rm FA}=\frac{1}{\sqrt{3}}\sqrt{\frac{(\lambda_{1}-
    \lambda_{3})^{2}+(\lambda_{2}-\lambda_{3})^{2}+(\lambda_{1}
    -\lambda_{2})^{2}}{\lambda_{1}^{2}+\lambda_{2}^{2}+
    \lambda_{3}^{2}}},
\label{equ:fa}
\end{equation}

where $\lambda_{1},\lambda_{2},\lambda_{3}$ are the eigenvalues of the Hessian
matrix of the density field. According to its definition, $\rm FA$ ranges between 0 and 1. In general, a lower FA means an equal collapse along the three eigenvectors, which corresponds to a cluster environment. A higher $\rm FA$ means that the collapse is highly anisotropic and being significant along one direction, usually corresponding to a sheet region. 

To understand the effect of FA on the spin-$\eee$ correlation, in Fig.~\ref{fig:fa} we first show the mass dependence of the $\rm FA$ parameters for haloes in different environments and redshifts. The dotted lines are for haloes in filaments and solid lines for haloes in clusters, while the  colours denote different redshifts. It is found that at a given redshift and environment, high mass haloes have lower $\rm FA$, and low-mass end haloes have higher $\rm FA$ but with a weak mass dependence. As pointed out by \cite{2013MNRAS.428.2489L}, this is a consequence of mass collapse around dark matter haloes, as a halo is more likely to grow when the mass on all directions of the LSS is strongly compressed. The figure also shows that $\rm FA$ is a strong function of environment and redshift. For haloes in clusters, the collapse along all the three eigenvector directions have happened, so the anisotropy is lower. In filament regions, instead, the collapse is mainly on the $\e$ and $\ee$ directions, so the anisotropy is systematically higher. We can see that the FA is also lower at low redshift. This is expected as mass collapse progressively proceeds to all directions with increasing time, as predicted by the Zel'dovich theory \citep[][]{1970A&A.....5...84Z}.

\begin{figure}
\centerline{\epsfig{figure=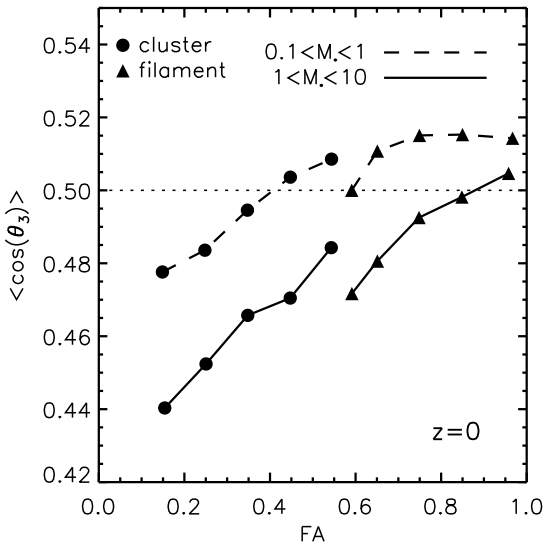,width=0.45\textwidth}}
\caption{
The mean $\rm cos(\theta_3)$ of cluster haloes
(filled circles) and filament haloes (filled upward triangles) as a function of 
$\rm FA$. Here we only show results at $z=0$ for halos in two mass bins.}
\label{fig:fa_filament_cluster}
\end{figure}

In Fig.~\ref{fig:fa_filament_cluster} we show the average alignment signal of spin-$\eee$ at $z=0$ as a function of $\rm FA$, for haloes in two mass bins and two environments, respectively. The general trend is that the halo spin tends to be perpendicular to $\eee$ with decreasing $\rm FA$. As discussed before, this is expected as a lower $\rm FA$ indicates that the mass along the slowest collapse direction $\eee$ also begins, so the mass is basically accreted along $\eee$ and results in a perpendicular signal. For the same reason, haloes in cluster regions tend to have a lower alignment between their spin and the LSS. However, it is interesting to see that for haloes with given $\rm FA$ and given environment, there is still a mass dependence such that high-mass haloes have a stronger mis-alignment between their spin and $\eee$. This can be explained by the halo formation and the migration time when the halo environment changes. \cite{2017MNRAS.468L.123W} have shown that in filament environments, massive haloes form after they enter into filaments, but low-mass haloes form before they enter a filament. The later significant mass accretion in massive haloes along with filament will lead to a stronger mis-alignment between the halo spin and the filament direction $\eee$. The results in Fig.~\ref{fig:fa_filament_cluster} show that there is no one single parameter which can fully determine the degree of alignment between the halo spin and the LSS, and the alignment is dependent on halo mass, environment and the anisotropy of the tidal field.


\section{Conclusions and discussion}
\label{sec:con}
In this paper, we study the correlation between halo spin and the slowest collapse direction, $\eee$, of the large scale structure. We investigate how the spin-$\eee$ correlation is built up over time, its relation to the anisotropic mass accretion history and the degree of anisotropic collapse on large scales. The main results are summarized as below,

\begin{itemize}

\item Similar to previous studies on the halo spin-$\eee$ correlation in filament and cluster environments separately, we find that, regardless of the environment, the halo spin-$\eee$ relation also has a mass dependence. For low-mass haloes, their spin tends to be parallel to $\eee$, and high mass haloes have spin perpendicular with $\eee$. The transition mass for the flip of the correlation is around $10^{12}M_{\odot}$, with a slight dependence on the smoothing length. We also find that  at a given halo mass, normalized by the characteristic mass of halo collapsing at a given redshift, halo spin is more likely to be parallel to $\eee$ at high redshift, but perpendicular to $\eee$ at lower redshift. 

\item By tracing the main progenitor of halo back in time, we study the build up of the halo spin-$\eee$ relation at different redshifts. we find that at early times the spin of progenitor haloes tend to be parallel to $\eee$, but the final correlation at $z=0$ depends on the mass growth. For massive halo, the mass collapse on large scale is significant and most of the mass accretion is along with the $\eee$ direction at low redshift. Under the assumption that the orbital angular momentum is transferred into the halo spin, it is naturally expected that mass accretion along $\eee$ will lead to a spin perpendicular to $\eee$. Massive haloes have quick evolution with redshift than low-mass ones as they migrate into their current environment earlier, as explained in \cite{2017MNRAS.468L.123W}.

\item The degree of anisotropic collapse along different directions of the tidal field can be described by the parameter $fractional$ $anisotropy$ (FA) proposed by \cite{2013MNRAS.428.2489L}. A higher FA means that the collapse on large scale is highly anisotropic, as in the sheet environment, and a lower FA indicates that near equal collapse has happened along all the three directions on large scale, which usually corresponds to a cluster environment. We find that at given halo mass and environment, the spin of halo with lower FA is more mis-aligned with the $\eee$. This is because in low FA region matter collapse has happened along the $\eee$ direction, leading to a halo spin inclined to be perpendicular to $\eee$. However, our results also show that FA is not the only parameter which can fully specify the spin-$\eee$ relation, but with additional dependence on halo mass and environment. 

\end{itemize}

Most previous works studied the spin-LSS correlations for haloes in different environments separately. Our study use haloes in different environments as a whole and the definition of LSS is consistent across different environments, thus the conclusion on the spin-LSS relation in this paper is more general. We trace the evolution of the spin-LSS correlation and find that the spin of massive haloes is not born to be mis-aligned (or perpendicular) with the LSS, but it is a consequence of the evolution of halo mass and environment. We also find that the correlation is dependent on the degree of anisotropy of the mass distribution on large scales.

Our results are consistent with the mass flow within the cosmic web as illustrated in \cite{1991QJRAS..32...85I} and \cite{2014MNRAS.441.2923C}. At early times, most haloes are formed in highly anisotropic environments, such as sheets, the mass is flowing from voids into the sheet planes along the fast collapsing direction, so the resulted halo spin is parallel to the slowest collapsing direction which lies in the sheet plane. As time goes by, haloes flow into filaments and clusters. Usually massive haloes are migrated into these environments earlier, so most of their mass growth is occurring in filaments or clusters in which the mass flow is mainly along the slowest collapsing direction, then the orbital angular momentum of the accreted mass will be transferred into halo spin, which will then be more likely to be perpendicular with the slowest collapsing direction of the LSS. Such a formation time and transition time dependence are also presented in our previous work by \cite{2017MNRAS.468L.123W}.

However, due to the complicated interplay between halo mass growth, environment changing and the anisotropy of the large-scale tidal field, we do not find a universal spin-$\eee$ correlation,  but being a function of redshift  and the anisotropy of the LSS.  For the first time, we find that the spin-$\eee$ correlation is a strong function of the anisotropy of the large-scale environment using N-body simulations. Such an environmental dependence of spin-filament correlation has been reported in observations by \cite{2010MNRAS.408..897J} in which they found that the spin of spiral galaxies is more perpendicular to the spine of the parent filament in the less dense environment. This is qualitatively in agreement with our findings in the paper. It is deserved to investigate the dependence of spin-filament alignment on the anisotropy of the large-scale environment using galaxy surveys in more detail and such a study will shed more light on our understanding of galaxy formation in the cosmic web.

\section{Acknowledgments}
We thank the referee, Miguel Arag{\'o}n-Calvo, for careful reading and constructive suggestions which improve the presentation of our paper. We also thank Liang Wang, Simon H. Wang and Emanuele Contini for careful reading and editing of the manuscript. We thank Noam Libeskind, John Peacock, Yanchuan Cai, Cautun Marius, Shi
Shao and Mark Neyrinck for discussions.  The work   is supported  by the  973
program  (No.  2015CB857003, No.  2013CB834900), the NSFC (No.  11333008), NSF of Jiangsu Province (No. BK20140050). The simulations are run on the supercomputing center of PMO and CAS.





\bibliographystyle{mn2e}
\bibliography{biblio}





\bsp	
\label{lastpage}
\end{document}